\documentclass{emulateapj}

\usepackage{amsmath}
\usepackage{amsfonts}
\usepackage{amssymb}
\usepackage{appendix}
\usepackage{bm}
\usepackage{cases}
\usepackage{float}
\usepackage{graphicx}
\usepackage{latexsym}
\usepackage{longtable}
\usepackage{multirow}
\usepackage{threeparttable}
\usepackage{txfonts}
\usepackage{tipa}

\shorttitle{Halo Substructures}
\shortauthors{Hefan Li et al.}

\begin{document}

\title{Two substructures in the nearby stellar halo found in Gaia and RAVE}

\author{Hefan Li\altaffilmark{1}, Cuihua Du\altaffilmark{2}, Yanbin Yang\altaffilmark{3},
Heidi Jo Newberg\altaffilmark{4}, Jun Ma\altaffilmark{5,2}, Jianrong Shi\altaffilmark{5,2}, Yunsong Piao\altaffilmark{1}}

\affil{$^{1}$School of Physical Sciences, University of Chinese Academy of Sciences, Beijing 100049, P. R. China\\
$^{2}$College of Astronomy and Space Sciences, University of Chinese Academy of Sciences, Beijing 100049, P.R. China; ducuihua@ucas.ac.cn\\
$^{3}$GEPI, Observatoire de Paris, PSL Research University, CNRS, Place Jules Janssen, F-92190 Meudon, France\\
$^{4}$Department of Physics, Applied Physics and Astronomy, Rensselaer Polytechnic Institute, Troy, NY 12180, USA\\
$^{5}$Key Laboratory of Optical Astronomy, National Astronomical Observatories, Chinese Academy of Sciences, Beijing 100012, China\\
}

\begin{abstract}

\par We use the second Gaia data release (Gaia DR2), combined with RAVE spectroscopic surveys, to identify the substructures in the nearby stellar halo. We select 3,845 halo stars kinematically and chemically, and determine their density distribution in energy and angular momentum space. To select the substructures from overdensities, we reshuffle the velocities and estimate their significance. Two statistically significant substructures, GR-1 and GR-2, are identified. GR-1 has a high binding energy and small $z$-angular momentum. GR-2 is metal-rich but retrograde. They are both new substructure, and may be accretion debris of dwarf galaxies. 

\end{abstract}

\keywords{Galaxy: kinematics and dynamics - Galaxy: halo - Solar neighborhood}

\section{Introduction}

\par According to the standard Lambda cold dark matter ($\Lambda$CDM) model,  smaller scale objects could merge into larger galaxies because of gravity \citep{White78}. In this merging process, dwarf galaxies and their tidal debris are dispersed into stellar halo \citep{Searle78, Freeman02}. These substructures arising from accretion event provide important clues for the study of galaxy formation. Although losing their spatial correlation, they still are linked by their phase space properties \citep{Helmi99,Bullock05}. \citet{Grillmair16} summarized the methods how to identify halo streams and provide the information of most streams and clouds discovered before 2015.

\par With the development of modern sky surveys, such as Sloan Digital Sky Survey \citep[SDSS;][]{York00}, Hipparcos \citep{Leeuwen07}, Large Sky Area Multi-Object Fiber Spectroscopic Telescope \citep[LAMOST;][]{Cui12,Deng12,Zhao12}, Radial Velocity Experiment \citep[RAVE;][]{Steinmetz06} and Gaia mission \citep{Gaia18}, more and more substructures are being discovered, such as Sagittarius dwarf tidal stream \citep{Ibata94,Majewski03,Belokurov06}, Orphan Stream \citep{Grillmair06,Belokurov07,Newberg10} and Virgo Stellar Stream \citep{Duffau06,Newberg07,Duffau14}.

\par Gaia DR2 \citep{Gaia18} provides positions, proper motions, and parallaxes for more than one billion stars. The precision and accuracy of these astrometric and photometric properties are unprecedented. Especially, Gaia DR2 also provides radial velocities for over 7 million stars. This allows us to get the full 6D phase space information of the star. Using metal-rich halo stars with 7D measurements supplied by Gaia and SDSS, \citet{Belokurov18} find an elongated structure in velocity space which they refer to as ``Gaia Sausage''. They show that it could be deposited in a major accretion event by a massive satellite. Similarly, \citet{Helmi18} show that the inner halo is dominated by the streams with slightly retrograde and elongated trajectories. These streams originate from accretion debris of a massive object that they call ``Gaia-Enceladus''. The stars of this structure have relatively low [$\alpha$/Fe] and a large spread in [Fe/H]. This also suggests that they formed in a relatively massive system. According to \citet{Helmi18}, the accretion of ``Gaia-Enceladus'' heats the thick disk of the Galaxy.

\par Radial Velocity Experiment Data Release 5 (RAVE DR5) contains 520,781 spectra of 457,588 unique stars in the magnitude range $9\ <\ I\ <\ 12$ \citep{Kunder17}. The medium-resolution spectra (R $\sim$ 7500) ensures the precision of radial velocity, atmospheric parameters and chemical abundances. Using Gaia DR2 and RAVE DR5, we can get kinematics and chemistry of stars. This can help us understand the early accumulation history of the Galaxy.

\par In this study we introduce the observational data from Gaia and RAVE in Section \ref{data}. Then we assemble a sample of halo stars selected from kinematic and chemical properties of stars. We discuss the method to identify substructures and estimate their significance in Section \ref{analysis}. Finally, we summarize our work in Section \ref{summary}.

\section{DATA AND METHODS}
\label{data}

\par Using quasars and validation solutions, \citet{Lindegren18} determined the systematic error of Gaia DR2 parallaxes to be $-0.029$ mas. \citet{huang18} determined that the RV zero-points (RVZPs) of Gaia$-$RVS and RAVE DR5 are 0.36 km/s and 0.17 km/s based on APOGEE \citep{Majewski17} data. So we correct the values of radial velocity of our sample stars for RVZP using their determinations. We cross-match Gaia and RAVE catalogs and obtain 512,971 stars in common. If the radial velocity of a star is provided by both Gaia and RAVE, its radial velocity and error are given by:
\begin{equation*}
rv = \frac{rv_G\,\sigma_R^2 + rv_R\,\sigma_G^2}{\sigma_G^2 + \sigma_R^2},\ \sigma_{rv}^2 = \frac{\sigma_R^2\,\sigma_G^2}{\sigma_G^2 + \sigma_R^2}
\end{equation*}
where $G$ and $R$ represent Gaia and RAVE, respectively. We consider only stars with relative parallax error $\leq 20\%$, signal-to-noise ration of RAVE (SNR) $\geq 20$ and radial velocity error $\sigma_{rv} \leq$ 10 km s$^{-1}$. We also use the selection criterion $|rv_G - rv_R| \leq max (10,\sigma_R,\sigma_G)$ km/s to ensure radial velocity reliability. Finally, the data set contains 372,321 stars.

\par We use the full Bayesian method to estimate distances and velocities \citep{Luri18}:
\begin{equation}
P(\bm{\theta}\ |\ \bm{x}) \propto \exp [-\frac{1}{2} (\bm{x} - \bm{m(\theta)})^\mathrm{T} C_x^{-1} (\bm{x} - \bm{m(\theta)})]\ P(d\ |\ L)
\end{equation}
where $\bm{\theta} = (d,\ v_{\alpha},\ v_{\delta},\ v_r)^\mathrm{T}$, $\bm{x} = (\varpi,\ \mu_{\alpha^*},\ \mu_{\delta},\ rv)^\mathrm{T}$, $\bm{m} = (1/d,\ v_{\alpha}/kd, \ v_{\delta}/kd,\ v_r)^\mathrm{T}$, $k$ = 4.74047 and $C_x$ is covariance matrix:
$$
\begin{pmatrix}
\sigma_{\varpi}^{2} & \rho_\varpi^{\mu_{\alpha^{*}}} \sigma_{\varpi} \sigma_{\mu_{\alpha^*}}  & \rho_\varpi^{\mu_{\delta}} \sigma_{\varpi} \sigma_{\mu_{\delta}} & \rho_\varpi^{rv} \sigma_{\varpi} \sigma_{rv} \\
\rho_\varpi^{\mu_{\alpha^{*}}} \sigma_{\varpi} \sigma_{\mu_{\alpha^{*}}} & \sigma_{\mu_{\alpha^{*}}}^{2} & \rho_{\mu_{\alpha^{*}}}^{\mu_{\delta}} \sigma_{\mu_{\alpha^{*}}}\sigma_{\mu_{\delta}} & \rho_{\mu_{\alpha^{*}}}^{rv} \sigma_{\mu_{\alpha^{*}}} \sigma_{rv} \\
\rho_\varpi^{\mu_{\delta}} \sigma_{\varpi} \sigma_{\mu_{\delta}} & \rho_{\mu_{\alpha^{*}}}^{\mu_{\delta}} \sigma_{\mu_{\alpha^{*}}}\sigma_{\mu_{\delta}} & \sigma_{\mu_{\delta}}^{2} & \rho_{\mu_{\delta}}^{rv} \sigma_{\mu_{\delta}} \sigma_{rv} \\
\rho_\varpi^{rv} \sigma_{\varpi} \sigma_{rv} & \rho_{\mu_{\alpha^{*}}}^{rv} \sigma_{\mu_{\alpha^{*}}} \sigma_{rv} & \rho_{\mu_{\delta}}^{rv} \sigma_{\mu_{\delta}} \sigma_{rv} & \sigma_{rv}^{2}.
\end{pmatrix}
$$
Prior of distance is the exponentially decreasing space density prior introduced in \citet{Bailer18}:
\begin{equation}
P(d\ |\ L(l, b)) \propto d^2 \exp (-d/L(l, b))
\end{equation}
where $L(l, b)$ is length scale, depending on Galactic longitude and latitude. We assume that the prior of $v_{\alpha}, v_{\delta}, v_r$ are uniform and they are only non-zero between 0 and $v_\mathrm{max}$ = 750 km s$^{-1}$, which is determined by experience. We choose most probable value of $d, v_{\alpha}, v_{\delta}, v_r$ as estimation of distances and velocities.

\par We use distances and Galactic coordinates $(l, b)$ to obtain Galactocentric Cartesian coordinates $(x, y, z)$ as follows \citep{Juric08}:
\begin{equation}
\begin{split}
&x = R_{\odot} - d \cos(b) \cos(l)\\
&y = - d \cos(b) \sin(l)\\
&z = d \sin(b) + z_{\odot}\\
\end{split}
\end{equation}
where $R_{\odot}$ = 8.2 kpc is galactocentric distance of the Sun \citep{Bland16}, $z_{\odot}$ = 25 pc is the solar offset from local disk mid-plane \citep{Juric08}, $d$ is distance and ($l$, $b$) are the Galactic coordinates. Galactic space-velocity ($U$, $V$, $W$) can be calculated from $d,\ v_{\alpha},\ v_{\delta},\ v_r$ \citep{Johnson87}. We correct the velocities based on the solar peculiar motion $(U_{\odot}, V_{\odot}, W_{\odot}) = (10., 11., 7.)$ km s$^{-1}$ \citep{Tian15, Bland16} and the LSR velocity $V_\mathrm{{LSR}}$ = 232.8 km s$^{-1}$ \citep{McMillan17}. $x$ and $z$ are positive in the directions of the Sun, and the North Galactic Pole (NGP), respectively. The $y$ axis is defined so as to keep the system right handed. $U$, $V$ and $W$ point toward the Galactic center, Galactic rotation, and the North Galactic Pole (NGP), respectively.

\begin{figure}
	\centering
	\includegraphics[width=1.0\hsize]{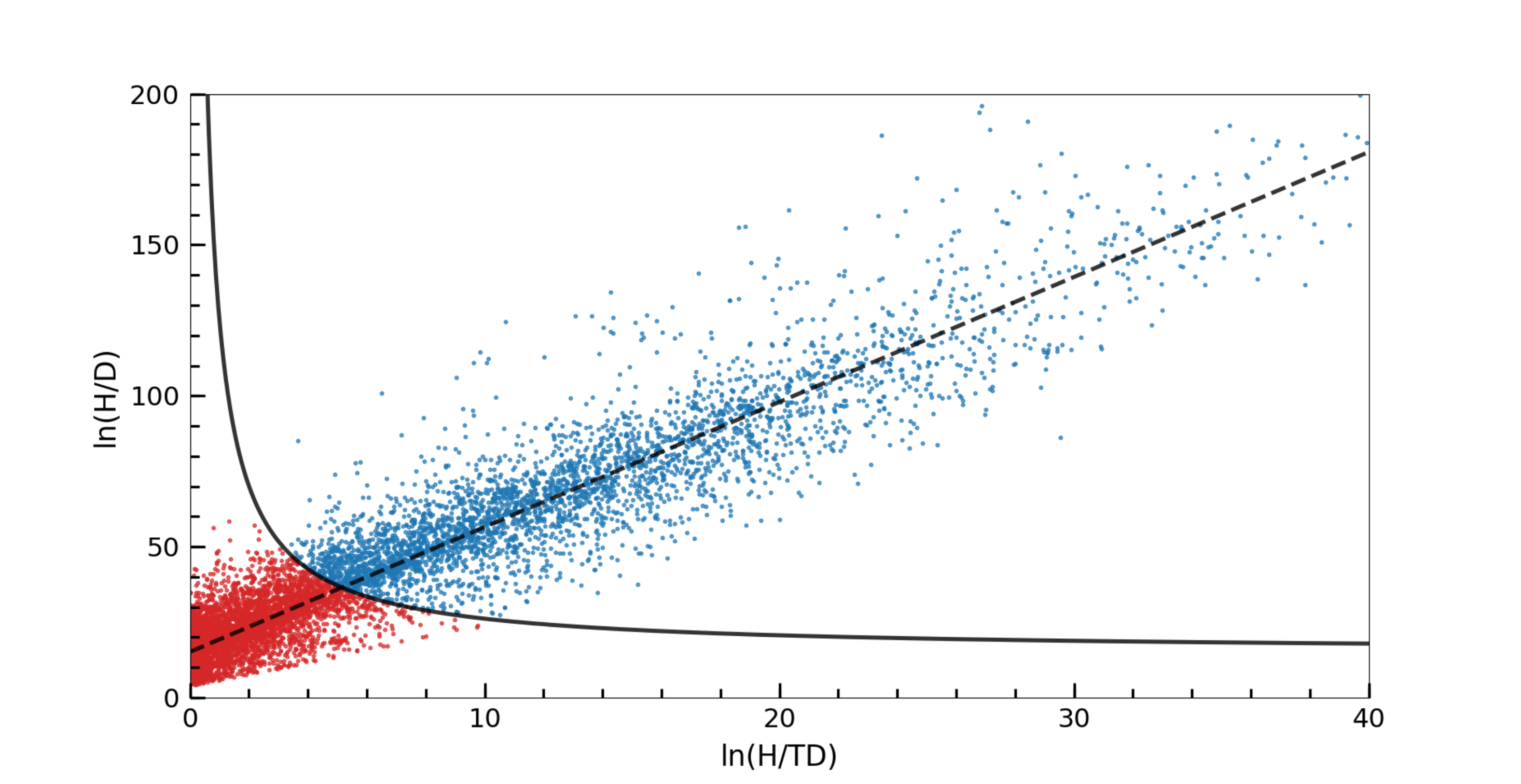}
	\caption{Distribution of $ln(\mathrm{H/D})$ vs $ln(\mathrm{H/TD})$ for stars with H$>$D and H$>$TD. The black dash line is the linear function for fitting $ln(\mathrm{H/TD})$ and $ln(\mathrm{H/D})$ using least square method. Halo stars (blue dots) are defined as having $ln(\mathrm{H/TD})*(ln(\mathrm{H/D})-b) >110$, and the dividing line is shown in solid black.}
	\label{HDTD}
\end{figure}

\begin{figure}
	\centering
	\includegraphics[width=1.0\hsize]{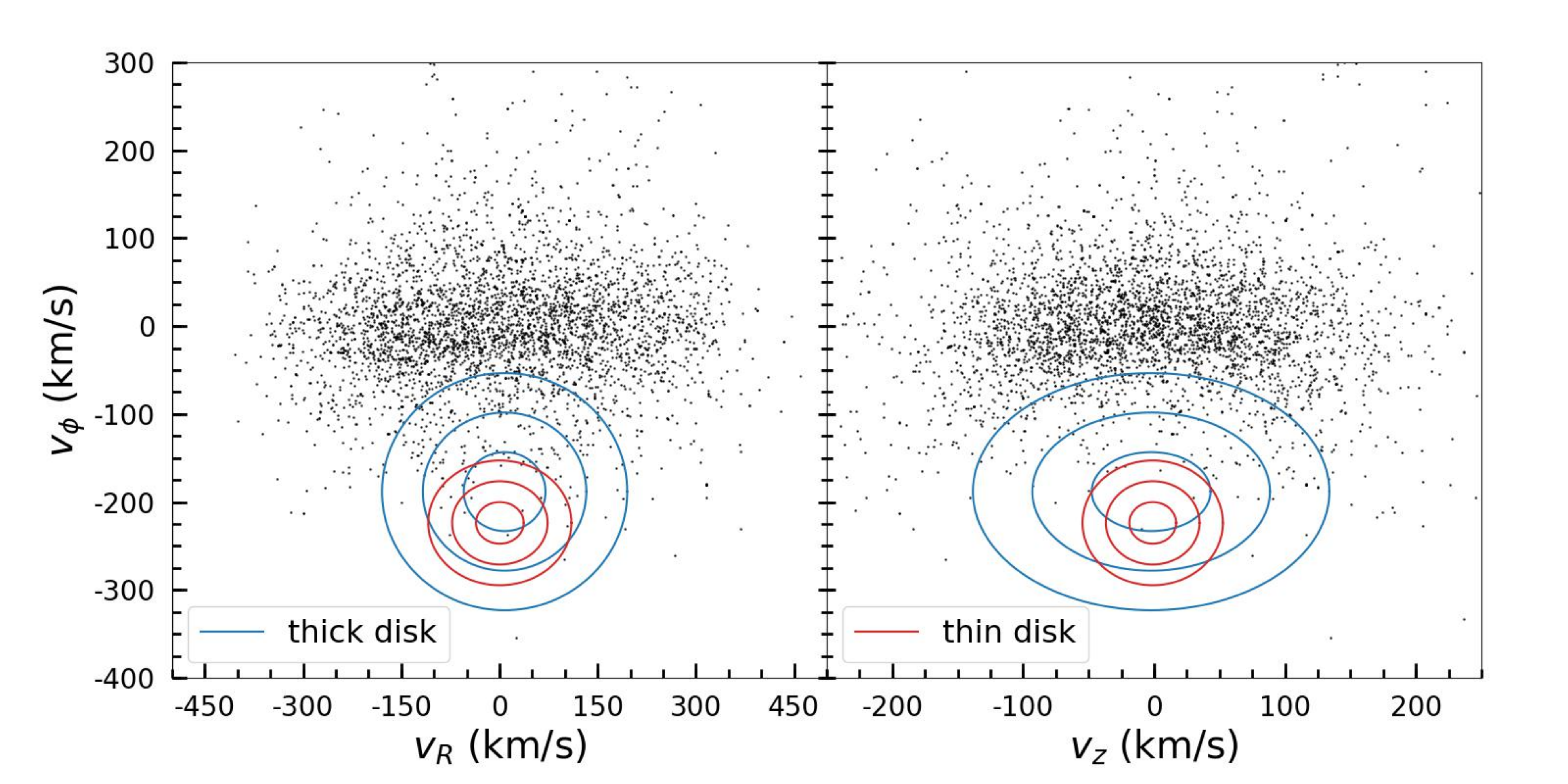}
	\caption{Velocities distribution of our halo sample (black dots) and probability contours of two Gaussian components (thin \& thick disk). The red (blue) lines show the 1 $\sigma$, 2 $\sigma$ and 3 $\sigma$ thin (thick) disk contour from inside to outside.}
	\label{halo}
\end{figure}

\begin{figure}
	\centering
	\includegraphics[width=1.0\hsize]{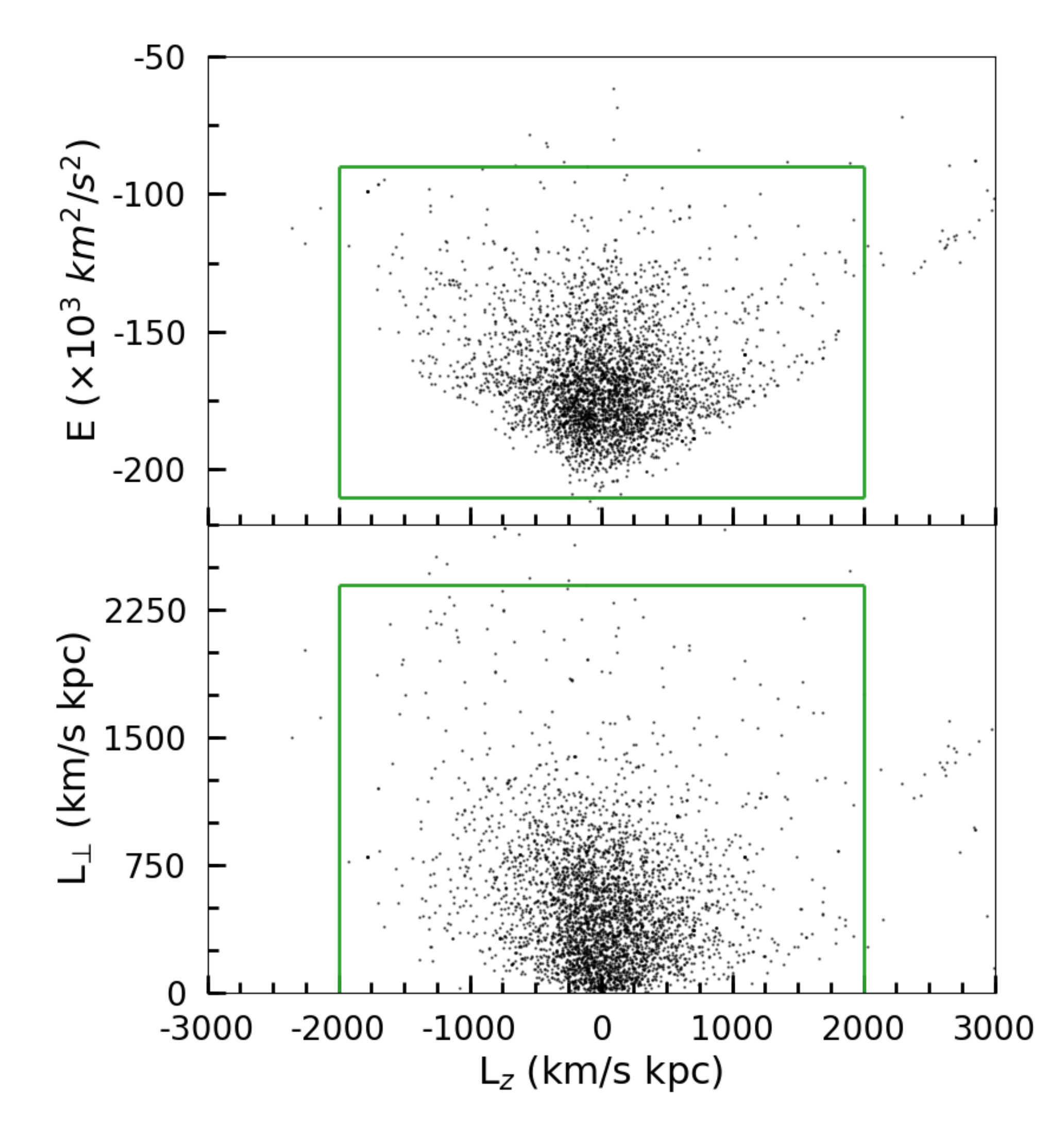}
	\caption{Distribution of Energy vs $L_z$ (top panel), and $L_{\bot}$ vs $L_z$ (bottom panel) for the halo sample that was selected in velocity and metallicity from the stars that were in common to both Gaia and RAVE. The green rectangle shows range of Figure \ref{kde}.}
	\label{ELL}
\end{figure}

\par Halo stars are commonly selected based on metallicity or kinematics. They are considered to be more metal-poor than the disk, and typical metallicity of the inner halo is from [Fe/H]$=-1.4$ \citep[e.g.,][]{An15,Zuo17,Gu15,Gu19} to [Fe/H] $= -1.6$ \citep[e.g.,][]{Allende Prieto06, Carollo07, Carollo10}. The criterion of kinematic selection is usually $\left| \textbf{V} - \textbf{V}_{\mathrm{LSR}} \right| > \textbf{V}_{\mathrm{cut}}$ in Toomre diagram. 

\par In addition, \citet{Posti18} select halo stars based on dynamic method. They compute the probability that a star belongs to a given component based on star's actions and a self-consistent total gravitational potential. They also compare their method with two other methods: kinematic selection and metallicity selection. The former is based on Toomre diagram we mentioned above. The latter first applied selection criteria [Fe/H] $\leqslant -1$ dex and then used two-Gaussian decomposition to further select in the velocity distribution. Only kinematically selected halo shows a strong bias in the velocity distribution, and it is difficult to identify halo stars in the velocity space dominated by the disk stars.
However, sample selected by metallicity does not contain stars with [Fe/H] $> -1$ dex, so that halo stars with richer metallicity are discarded by the method.
But in dynamic and kinematic method, the metallicity distribution is centered on [Fe/H] $\sim-0.5$ dex and extending out to [Fe/H] $\sim 0$. 

\par For a given star, \citet{Bensby03} calculate the probability that it belongs to the thin disk, the thick disk, and the stellar halo by assuming that the Galactic space velocities ($U,V,W$) of each population have Gaussian distributions.  

\par On this basis, we use cylindrical velocities and spherical velocities instead of the Cartesian velocities and assume that the metallicity has also Gaussian distribution. Halo stars are much less than disk stars, so the probability that stars belong to the disk and the halo will be calculated separately.

\par In this work, we adopt the mean rotation of halo stars $\bar{v}_{\phi}$=$-38.8$ km s $^{-1}$ according to $\bar{v}_{\phi}\ =\ -(V_\mathrm{LSR}\ +\ V_{\odot})\ +\ \Delta v_\phi$, 
$\Delta v_\phi$ is the difference between $\bar{v}_\phi$ of halo stars  and $v_{\phi,\ \odot}$,
$\Delta v_\phi=205$ km s$^{-1}$ is derived from the study of \citet{Bond10}  ($\bar{v}_\phi$ = $-$20 km s$^{-1}$ for their adoption $V_\mathrm{LSR}$ = 220 km s$^{-1}$ and $V_{\odot}$ = 5 km s$^{-1}$).
Here we adopt recent $V_\mathrm{LSR}$ = 232.8 km s$^{-1}$\citep{McMillan17}, $V_{\odot}$ = 11 km s$^{-1}$\citep{Bland16}.  The velocity ellipsoid in spherical coordinates is $(\sigma_{r}, \sigma_{\theta}, \sigma_{\phi}) = (141, 75, 85)$ km s$^{-1}$ \citep{Bond10}. For the mean metallicty and dispersion of the halo stars, we use recent result $\overline{\mathrm{[Fe/H]}} = -1.43$ and $\sigma_{\mathrm{[Fe/H]}} = 0.36$ derived by \cite{Zuo17}.
So the probability that stars belong to halo (which we call H) can be given by a four-dimensional Gaussian distribution:

\begin{figure*}
	\centering
	\includegraphics[width=1\textwidth]{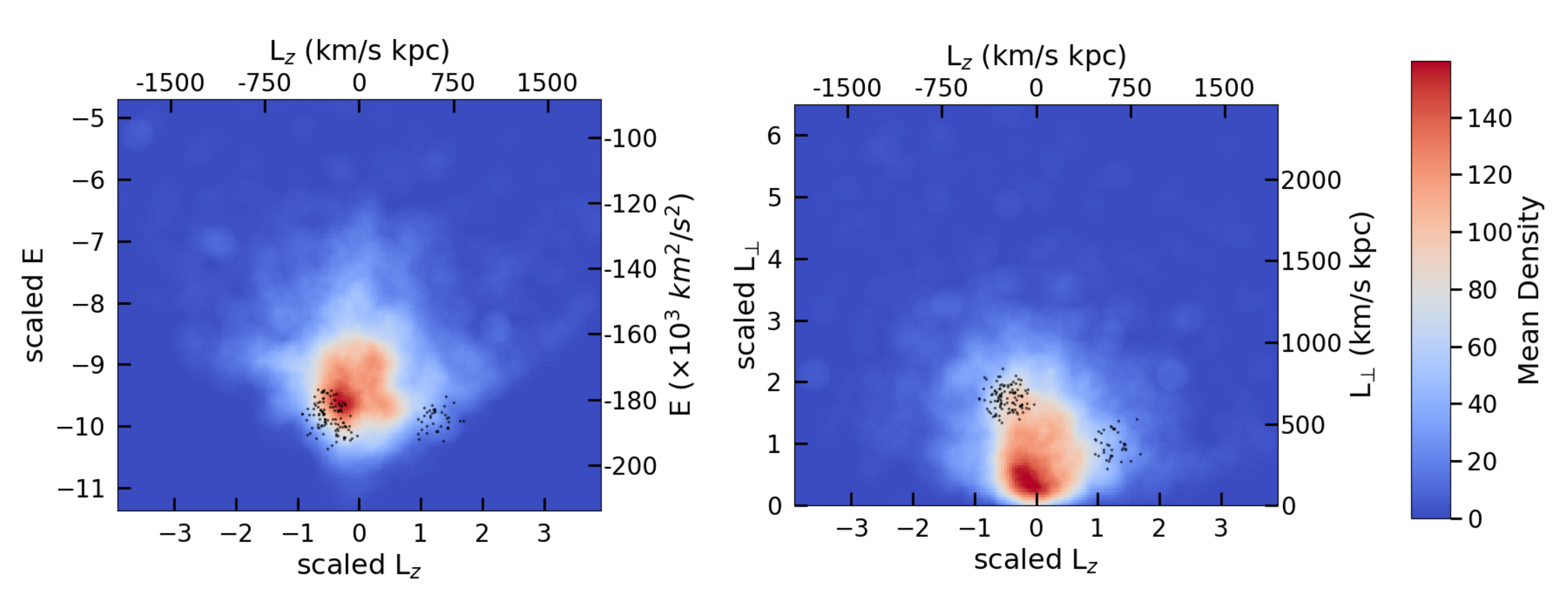}
	\caption{The mean of three-dimension ($E, L_z, L_{\bot}$) density $kde_{\mathrm{real}}$ along the $L_{\bot}$ (left panel) and $E$ (right panel). The black dots are the member stars of substructures.}
	\label{kde}
\end{figure*}

\begin{figure*}
	\centering
	\includegraphics[width=1\textwidth]{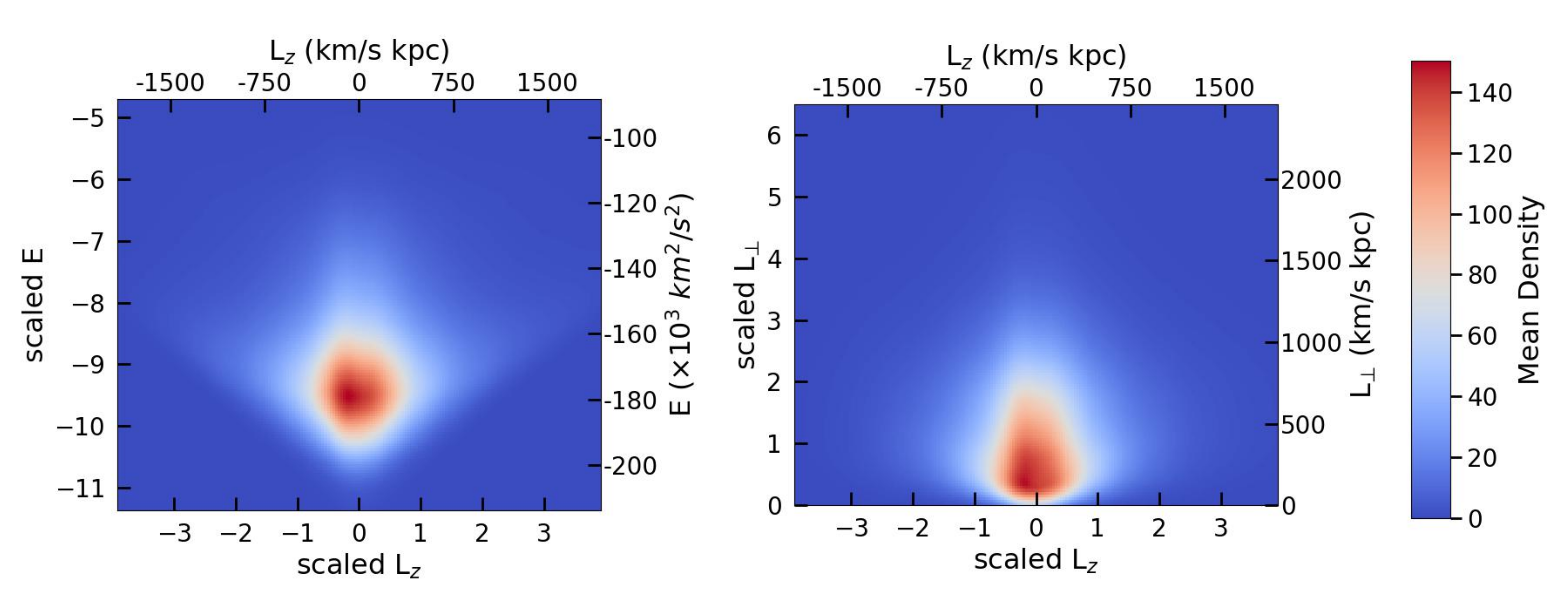}
	\caption{Similar to Figure \ref{kde}, but now we use average density of all 5000 randomized realizations of the data.}
	\label{lambda}
\end{figure*}

\begin{figure*}
	\centering
	\includegraphics[width=1\textwidth]{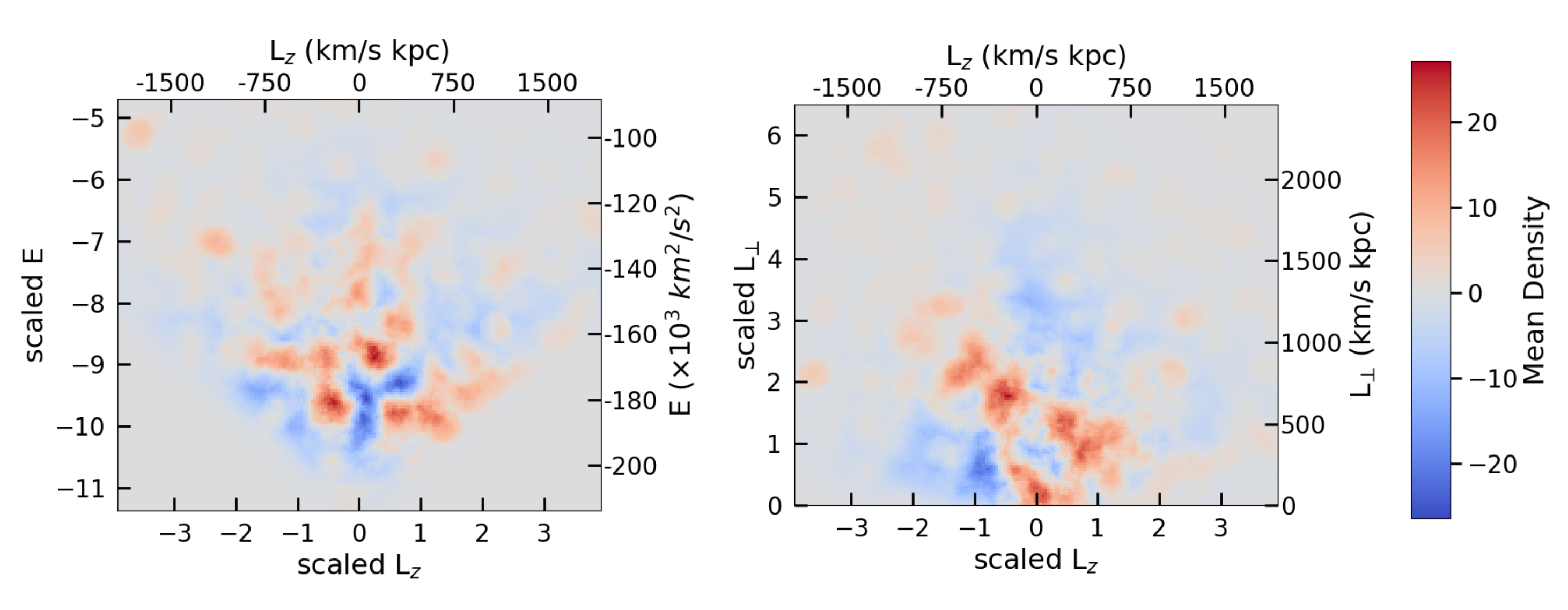}
	\caption{\textbf{Similar to Figure \ref{kde}, but now we use the difference between the kernel density of real data and the average density of all 5000 randomized realizations of the data.}}
	\label{diff}
\end{figure*}

\begin{equation}
\begin{split}
&f(v_{r},\, v_{\theta},\, v_{\phi},\, \mathrm{[Fe/H]}) =\\
&k \cdot
\exp \left( -\frac{v_r^2}{2 \sigma_r^2} - \frac{v_{\theta}^2}{2 \sigma_{\theta}^2} - \frac{(v_{\phi}-\bar{v}_{\phi})^2}{2 \sigma_{\phi}^2}
- \frac{(\mathrm{[Fe/H]}-\overline{\mathrm{[Fe/H]}})^2}{2 \sigma_{\mathrm{[Fe/H]}}^2} \right),\\
\end{split}
\label{gau}
\end{equation}
where $k = \frac{1}{(2\pi)^2\,\sigma_r\, \sigma_{\theta}\, \sigma_{\phi}\, \sigma_{\mathrm{[Fe/H]}}}$. For those stars with no metallicity parameters which is less than 10\% of our sample, we set their [Fe/H] to $-1$ in equation (\ref{gau}) to prevent the loss of halo stars.

\par Then we calculate the probability that stars belong to the thin disk and the thick disk (which we call D and TD). To minimize the influence of halo stars on disk, we here remove stars with $v_{\phi} > -40$ km s$^{-1}$ and [Fe/H] $< -1.4$. We use the $\textsc{sci-kit learn}$ package in $\textsc{python}$ \citep{Pedregosa12} to fit a two-component Gaussian Mixture Model to spherical velocities ($v_R$, $v_{\phi}$, $v_z$) and stellar metallicity ([Fe/H]). In this fitting, each component has its own diagonal covariance matrix. The velocity dispersions of the thin disk and thick disk is $(\sigma_{R}, \sigma_{\phi}, \sigma_{z}) = (36, 24, 18)$ km s$^{-1}$ and $(\sigma_{R}, \sigma_{\phi}, \sigma_{z}) = (62, 45, 45)$ km s$^{-1}$, respectively. The mean $v_{R}$ and $v_{z}$ are close to 0 km s$^{-1}$ for both thin and thick disk. The mean of $v_{\phi}$ is $-$224 km s$^{-1}$ for the thin disk and $-$188 km s$^{-1}$ for the thick disk. The mean metallicity and dispersion are $-0.11,\ 0.28$ for the thin disk and $-0.47,\ 0.32$ for the thick disk.

\par We use the least square to fit the function:
\begin{equation}
ln(\mathrm{H/D}) = a*ln(\mathrm{H/TD}) + b
\end{equation}
to stars with H$>$D and H$>$TD. The optimal values for the parameters are $a=4.14$ and $b=15.14$. Then we select stars with $ln(\mathrm{H/TD})*(ln(\mathrm{H/D})-b) >110$ and get a similar number of stars with $L_z > 0$ and stars with $L_z < 0$. We finally get 3,845 sample halo stars. The result is shown in Figure \ref{HDTD}. Figure \ref{halo} show their velocity distribution and two Gaussian components (thin \& thick disk) in cylindrical coordinates system.

\section{Analysis}
\label{analysis}

\citet{Helmi17} use the ``Integrals of Motion" space, which consists of energy and two components of the angular momentum. They study the distribution of stars in this space and identify several substructures that could potentially be related to merger events. \citet{Li19} also identify halo substructures in this space and they improve the method of estimating statistical significance. Here, we use the latter method to find substructures from our sample stars. 
 
\par First, we compute energy and angular momentum by adopting a Galaxy potential model \citep{McMillan17} with four components: the cold gas discs near the Galactic plane, the thin and thick stellar discs, a bulge and a dark-matter halo. Figure \ref{ELL} shows the distribution in energy versus $z$-angular momentum, $L_z$, in the upper panel, and $L_{\bot} = \sqrt{L^2_x+ L^2_y}$ versus $L_z$ on the lower panel for our sample stars. The green rectangle in this figure shows the populated region, $-2000 < L_z < 2000$ km s$^{-1}$ kpc, $-2.10 \times 10^5 < E < -0.9 \times 10^5\ \mathrm{km^2\ s^{-2}}$, and $L_{\bot} < 2400$ km s$^{-1}$ kpc. 

\par Then we scale the data to unit standard deviation and calculate the stellar density (which we call $kde_0$) in $E - L_z - L_{\bot}$ space using kernel density estimator in $\textsc{sci-kit learn}$. We choose the ``Tophat'' kernel and the optimal bandwidth of kernel is 0.28, which is determined by the cross-validation method \citep[e.g.][]{Weiss91}, also implemented in $\textsc{sci-kit learn}$. To show the three-dimensional density $kde_0(E , L_z , L_{\bot})$ in the plane, we take the arithmetic mean value of $kde_0(E, L_z, L_{\bot})$ along the specified axis as a two-dimensional density function. In Figure \ref{kde}, the left panel shows the mean value along the $L_{\bot}$ and the right panel shows the mean value along the $E$. Because of the characteristics of the ``Tophat'' kernel, the number of stars in a spherical region with a radius of bandwidth can be derived from the density at the center of the sphere:
\begin{equation}
N_0 (E , L_z , L_{\bot}) = \dfrac{4}{3} \pi r^3 kde_0 (E , L_z , L_{\bot})
\end{equation}
where $r=0.28$ is the bandwidth.

\begin{figure}[htb]
	\centering
	\includegraphics[width=0.5\textwidth]{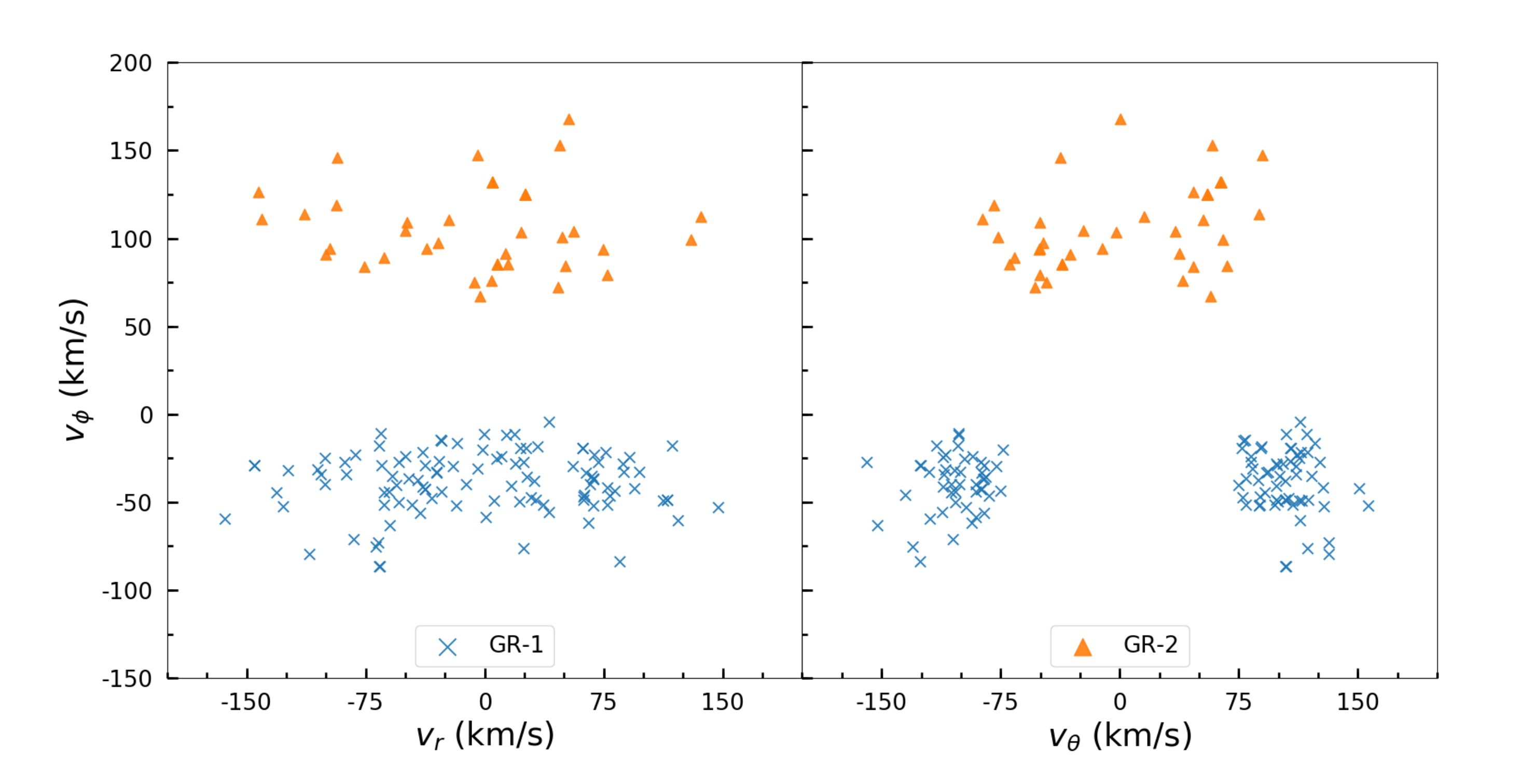}
	\caption{Spherical velocities distribution for stars comprising the identified substructures.}
	\label{struct}
\end{figure}

\par To identify the authentic overdensities instead of the relative peaks of density, we reshuffle each velocity component ($v_R, v_\theta, v_\phi$) to create 5,000 randomized datasets. This scramble the correspondence between stars and their velocity components and it means that each velocity component of a star may come from different stars. We recompute their $E, L_z, L_{\bot}$ and use the kernel density estimator with same parameters to get the densities, $kde_i (E, L_z, L_{\bot})$, and the numbers, $N_i (E, L_z, L_{\bot})$ just like we did with real data. We have generated 5000 random samples so $i=1,2,3,...,5000$. We use the Cumulative Distribution Function of Poisson distribution:
\begin{equation}
P(X<N_0) = \sum_{k=0}^{N_0-1} \dfrac{\lambda^k}{k!} \,e^{-\lambda}
\end{equation}
to estimate the significance. Parameter $\lambda$ is the mean of $N$ for 5000 randomized datasets:
\begin{equation}
\lambda(E, L_z, L_{\bot}) = \dfrac{1}{5000} \sum_{i=1}^{5000} N_i (E, L_z, L_{\bot}).
\end{equation}
Figure \ref{lambda} shows the average density of 5000 randomized datasets. We can see a V-shaped envelope similar to the density distribution of real data (Figure \ref{kde}), but it is more smooth. \textbf{Figure \ref{diff} shows the difference between the kernel density of real data and the average density of all 5000 randomized realizations of the data. There are several clearly overdensities.} We divide the $E - L_z - L_{\bot}$ space into smaller cube pixels that are about 0.04 on a side. These pixels are used as probes to scan the ``Integrals of Motion" space. We select pixels with significance $> 3.5 \sigma$ and $\lambda >3$. The clumps formed by adjacent pixels are part of one or more overdensities. We use $\textsc{sci-kit image}$ \citep{Walt14} implementation to remove very small clumps and connect two clumps that are close but not adjacent. We only consider overdensities bigger than 230 pixels. When we use the above method to identify overdensity from random data, we only find one overdensity in 100 randomized datasets.

\begin{figure}[htb]
	\centering
	\includegraphics[width=0.5\textwidth]{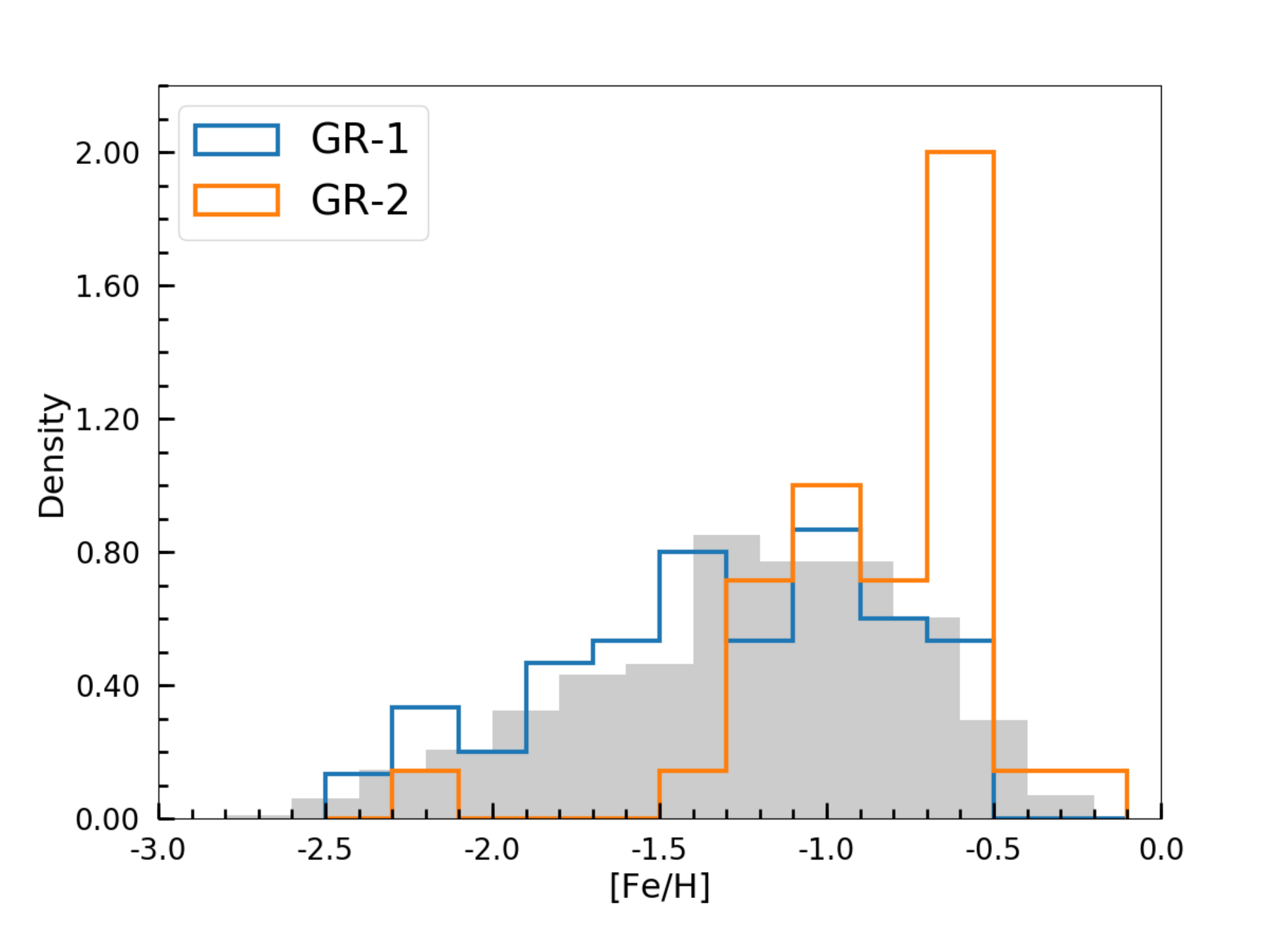}
	\caption{Metallicity distribution for the member stars of comprising the identified substructures. Halo sample is shown as gray histogram.}
	\label{feh}
\end{figure}

\begin{figure}[htb]
	\centering
	\includegraphics[width=0.5\textwidth]{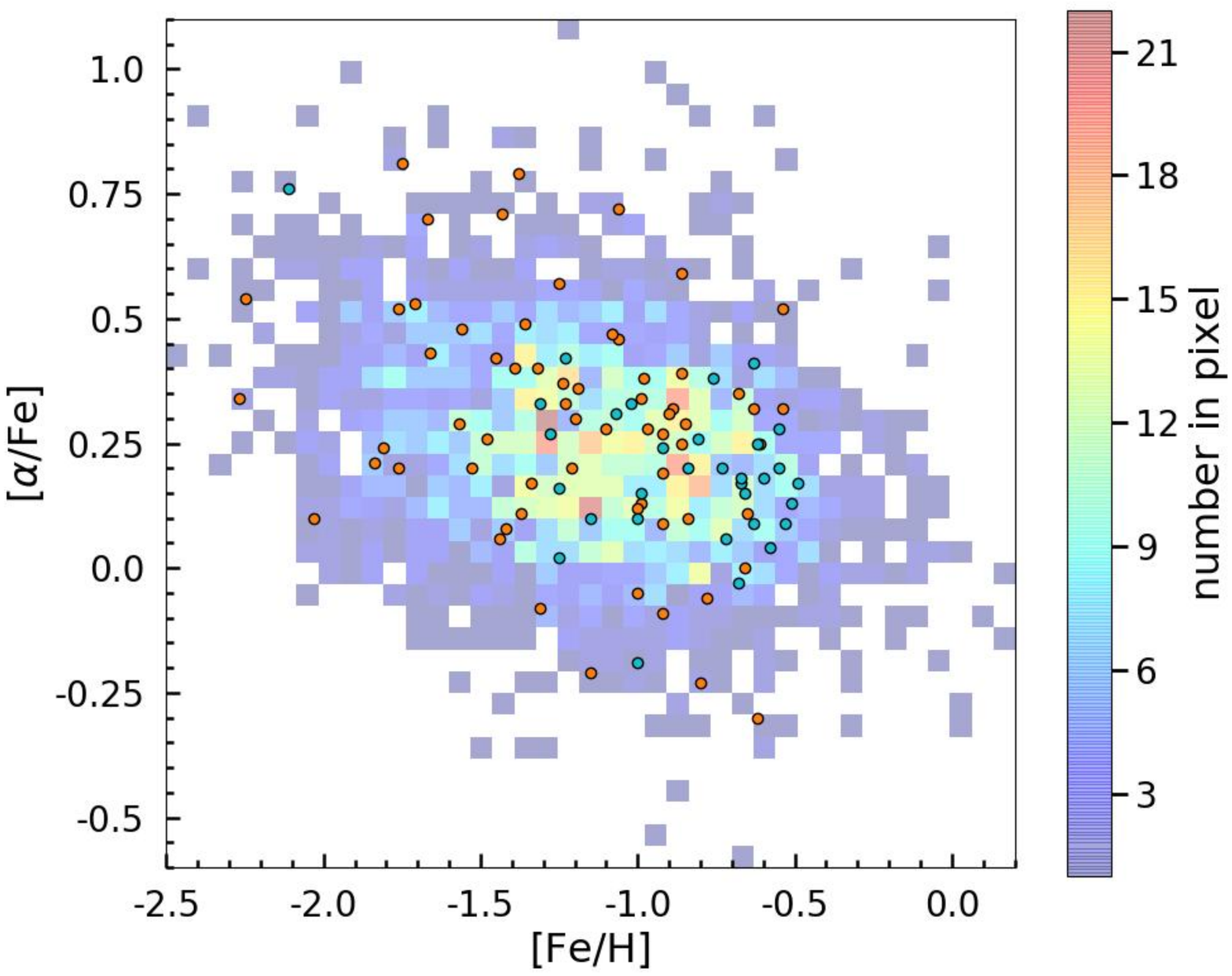}
	\caption{Chemical abundance distribution [$\alpha$/Fe] vs. [Fe/H]  for the member stars of comprising the identified substructures. The orange dottes represent `GR-1' and the green dottes represent `GR-2'. Halo stars are shown as background for comparison and the color coding corresponds to the number of halo stars in each pixel.}
	\label{alpha}
\end{figure}

\par We identify two overdensities in ``Integrals of Motion" space and the member stars are shown in Figure \ref{kde}. The significance of the complete overdensities is also estimated. These overdensities are identified as substructures in this study, and we label them GR-1 and GR-2. The information of member stars are listed in Table 1 and Table 2 in the Appendix. Velocities distribution of these member stars are shown in Figure \ref{struct}. These stars are separated in spherical velocities space. GR-1 shows distinct separate clumps in $v_{\phi} - v_{\theta}$ space, and the velocity distribution characteristic is similar to VelHel-7 of \citet{Helmi17} and GL-1, GL-2 of \citet{Li19}. Although VelHel-6 of \citet{Helmi17} is closer to GR-1 in ``Integrals of Motion" space, they have different characteristics in velocity space. Many more substructures have been reported by other authors \citep[e.g.,][]{Klement09,Zhao15}, but we have not found other substructures that are similar to ours. This may be due to different samples or methods. Figure \ref{feh} shows metallicity distribution for the member stars, GR-1 is more similar to GL-1 of \citet{Li19}. We consider that this could be related to multiple accumulation from the same dwarf galaxy, which need to further confirmation in the future. In addition, we can also see from Figure \ref{alpha} that most member stars in GR-2 are metal-rich and slightly $\alpha$-enriched, with a mean $\alpha$-abundance of $\overline{[\alpha/\rm Fe]}=+0.1$ dex, the member stars in GR-1 are metal-poor and $\alpha$-enriched with  a mean $\alpha$-abundance of $\overline{[\alpha/\rm Fe]}=+0.3$ dex. The mean and dispersion of energy and angular momentum of the above structures are listed in Table \ref{mean}. The \citet{McMillan17} potential model was used to recomputed the substructure energy found by other authors.

\begin{center}
	\begin{longtable}{lccc}
		\label{mean}\\
		\caption{The mean and dispersion of energy and angular momentum of substructures}\\
		\hline
		\hline
		Notation  &  E  &    $L_z$    &    $L_{\bot}$    \\
		& ($\times 10^3\ \mathrm{km^2\ s^{-2}}$)   &  (km s$^{-1}$ kpc) & (km s$^{-1}$ kpc) \\
		\hline
		GR-1     &  -185 $\pm$ 4 &   -237 $\pm$ 96 &   659 $\pm$ 70 \\
		GR-2     &  -186 $\pm$ 3 &    622 $\pm$ 87 &   375 $\pm$ 74 \\
		GL-1     &  -168 $\pm$ 3 &  -809 $\pm$ 161 &  978 $\pm$ 132 \\
		GL-2     &  -167 $\pm$ 1 &   -368 $\pm$ 87 &  1284 $\pm$ 57 \\
		VelHel-6 &  -176 $\pm$ 4 &  -542 $\pm$ 330 &  244 $\pm$ 117 \\
		VelHel-7 &  -174 $\pm$ 6 &  -919 $\pm$ 313 &  563 $\pm$ 162 \\
		\hline
	\end{longtable}
\end{center}

\section{Summary and discussion}
\label{summary}

\par We cross-match Gaia DR2 and RAVE DR5 data to get a sample of stars with full phase-space information. By combining kinematic and chemical properties of stars, we obtain 3,845 halo sample stars. We determine the density field of the sample halo stars in ``Integrals-of-Motion" space, defined by two components of the angular momentum, $L_{\bot}$ and $L_z$, and energy. In order to identify authentic substructures from several overdensities in this space, we estimate the statistical significance for these overdensities and identify two substructures, GR-1 and GR-2. The catalog of two substructures' members are given in the Appendix.

\par In this study, both two substructures are new. The velocity distribution characteristic of GR-1 in our study is similar to VelHel-7  of \citet{Helmi17} and GL-1, GL-2 of \citet{Li19}. But they are separated in ``Integrals of Motion''. The metallicity of GR-2 is close to the thick disk, but GR-2 has retrograde trajectories. Their origin may be related to the accretion events of dwarf galaxy, which need to be confirmed in the following studies. \citet{Bonaca17} selected halo stars from Gaia, RAVE and APOGEE data kinematically, and they shown half of their halo sample stars have [Fe/H] $>$ $-1$.   
The orbital directions of these metal-rich halo stars are preferentially prograde,  which is different from the intrinsically isotropic orbital distribution of the metal-poor halo stars. At the same time, they also found similar properties in their Latte cosmological simulation and  suggested that metal-rich halo stars in the solar neighborhood formed in situ within the Galactic disk. Other studies also  implied that some halo stars are preferentially on slight prograde orbits \citep[e.g.,][]{Deason17,Kafle17,Bird19}. While the discovery of GR-2  having retrograde orbit suggests that metal-rich stars may also originate from accretion. The history of the formation of the Galaxy may still remain in the dynamical and chemical properties of substructures. This can help us understand the early stages of the Galaxy.

\section{Acknowledgements}

\par We thank especially the referee and statistical editor for insightful comments and suggestions, which have improved the paper significantly. This work was supported by the National Natural Foundation of China (NSFC No. 11973042 and No. 11973052).  HJN acknowledges funding from NSF grant AST 16-15688. Funding for SDSS-III has been provided by the Alfred P. Sloan Foundation, the Participating Institutions, the National Science Foundation, and the U.S. Department of Energy Office of Science. 

\par This work has made use of data from the European Space Agency (ESA) mission {\it Gaia} (\url{https://www.cosmos.esa.int/gaia}), processed by the {\it Gaia} Data Processing and Analysis Consortium (DPAC, \url{https://www.cosmos.esa.int/web/gaia/dpac/consortium}). Funding for the DPAC has been provided by national institutions, in particular the institutions participating in the {\it Gaia} Multilateral Agreement.

\par Funding for RAVE has been provided by: the Australian Astronomical Observatory; the Leibniz-Institut fuer Astrophysik Potsdam (AIP); the Australian National University; the Australian Research Council; the French National Research Agency; the German Research Foundation (SPP 1177 and SFB 881); the European Research Council (ERC-StG 240271 Galactica); the Istituto Nazionale di Astrofisica at Padova; The Johns Hopkins University; the National Science Foundation of the USA (AST-0908326); the W. M. Keck foundation; the Macquarie University; the Netherlands Research School for Astronomy; the Natural Sciences and Engineering Research Council of Canada; the Slovenian Research Agency; the Swiss National Science Foundation; the Science \& Technology Facilities Council of the UK; Opticon; Strasbourg Observatory; and the Universities of Groningen, Heidelberg and Sydney. The RAVE web site is at https://www.rave-survey.org.

\newpage
\appendix  
\renewcommand{\appendixname}{Appendix~\Alph{section}}  
\section{Member stars of the new identified substructures}
\centering
\par Unique source identifiers in Gaia, positions, metallicities, space velocities and z-angular momentum for the member stars.

\begin{center}
\begin{longtable}{ccccccccccc}
\caption{GR-1, 4.57 $\sigma$, 99 stars}\\
\hline
\hline
source-id & l &      b &  [Fe/H] &     x &    y &     z &      U &      V &      W &     $L_z$     \\
& (deg) & (deg) & (dex) & (kpc) & (kpc) & (kpc) & (km s$^{-1}$) & (km s$^{-1}$) & (km s$^{-1}$) & (km s$^{-1}$ kpc) \\
\hline\\
 6098916889620300672 &  323.15 &  14.25 &  -1.37 &  5.8 &  1.8 &  0.8 &   -4.0 &   11.3 &  102.6 &  -72.8 \\
4715376531932531200 &  296.65 & -54.22 &      - &  7.6 &  1.1 & -1.7 &   44.2 &   39.7 &  -84.7 & -254.8 \\
6360190882785741952 &  314.12 & -29.00 &  -1.31 &  6.9 &  1.3 & -1.0 &  -34.7 &   45.7 &  -83.7 & -361.7 \\
6822643716784658560 &   39.70 & -54.34 &  -1.34 &  7.1 & -0.9 & -1.9 &  -10.7 &   41.5 &  105.5 & -286.8 \\
6273670931270685184 &  323.42 &  33.63 &  -0.99 &  7.0 &  0.9 &  1.0 &   59.2 &   31.6 &   84.7 & -167.9 \\
6333120597271443456 &  351.45 &  43.27 &  -0.99 &  6.6 &  0.2 &  1.5 &   11.8 &   21.8 & -120.0 & -141.6 \\
4705104619428366720 &  297.60 & -51.35 &  -1.24 &  7.7 &  1.0 & -1.4 &  -86.2 &   26.0 &   72.5 & -284.4 \\
6697060904572244352 &    4.08 & -30.38 &  -0.85 &  6.1 & -0.1 & -1.2 &  -39.3 &   20.0 & -117.5 & -116.3 \\
6419114462535500160 &  321.46 & -22.26 &  -0.78 &  6.8 &  1.1 & -0.7 &  -17.2 &   16.6 &  -78.7 & -132.1 \\
6194053026560559488 &  309.59 &  38.58 &   -1.0 &  7.0 &  1.4 &  1.5 &  -43.7 &   21.3 &   87.4 & -210.9 \\
6718953383671971584 &  359.73 & -19.67 &      - &  5.5 &  0.0 & -0.9 &   63.7 &   71.0 &  117.3 & -392.7 \\
6806331572732029568 &   21.09 & -35.16 &  -1.23 &  6.0 & -0.9 & -1.7 &   11.6 &   26.5 & -100.3 & -168.3 \\
6467385187658987904 &  337.95 & -35.23 &  -1.32 &  7.0 &  0.5 & -0.9 &  -79.6 &   34.2 &   81.8 & -278.0 \\
4353736643479056384 &   13.12 &  25.24 &  -1.84 &  3.9 & -1.0 &  2.1 & -138.2 &   79.2 &  -89.7 & -169.2 \\
4195113433837448064 &   31.42 & -17.88 &  -1.48 &  6.0 & -1.3 & -0.8 &  -61.6 &   77.2 &   83.9 & -380.7 \\
4905172506836913920 &  306.76 & -56.55 &  -1.71 &  7.4 &  1.1 & -2.1 &   27.3 &   48.6 &  103.4 & -328.3 \\
6469232749446813824 &  342.02 & -34.68 &  -1.92 &  7.3 &  0.3 & -0.6 &   28.4 &   43.6 &   90.0 & -310.4 \\
6846461582481204096 &   15.92 & -30.53 &  -2.18 &  6.9 & -0.4 & -0.7 &  113.9 &   26.2 &  116.6 & -222.9 \\
6674640385011413120 &  352.97 & -36.08 &  -1.21 &  3.9 &  0.5 & -3.2 & -138.8 &   27.3 &   65.7 & -179.7 \\
4342588209752172160 &    1.94 &  27.96 &      - &  4.1 & -0.1 &  2.2 &   39.5 &   78.2 & -168.0 & -329.7 \\
6740456223339677312 &    4.82 & -26.64 &  -0.92 &  4.5 & -0.3 & -1.9 &   88.1 &   23.0 &  171.4 & -130.6 \\
6446375551973492992 &  336.94 & -28.45 &  -2.25 &  6.3 &  0.8 & -1.1 &    3.0 &   44.7 &  106.9 & -278.2 \\
6415779334530809728 &  321.53 & -27.67 &  -1.76 &  6.3 &  1.5 & -1.2 &  111.1 &   52.1 &  -92.1 & -161.5 \\
6450293932239999744 &  330.69 & -36.96 &      - &  6.8 &  0.8 & -1.2 &  -98.0 &   32.1 &   59.7 & -295.9 \\
5813464395238590976 &  325.97 & -16.33 &  -0.89 &  4.6 &  2.5 & -1.3 &  -50.5 &   31.8 & -168.5 & -269.1 \\
6407668963103576064 &  329.03 & -48.49 &      - &  6.6 &  1.0 & -2.1 &    2.5 &   50.7 & -101.4 & -329.6 \\
4414394741476513536 &    0.09 &  42.57 &  -0.92 &  6.1 & -0.0 &  1.9 &   32.8 &   44.4 & -106.1 & -271.8 \\
6253774890413942912 &  346.87 &  28.95 &   -1.0 &  5.8 &  0.6 &  1.4 & -140.6 &   34.9 &  -74.2 & -281.0 \\
6580544764022363520 &  359.77 & -44.69 &      - &  5.6 &  0.0 & -2.6 &  -29.1 &   51.3 & -121.3 & -285.8 \\
6278779051839299328 &  335.20 &  33.52 &  -1.76 &  5.4 &  1.3 &  2.1 &   89.5 &   99.6 &   97.5 & -415.5 \\
4181613110374697344 &   23.20 & -18.28 &  -1.36 &  4.7 & -1.5 & -1.2 &  -81.7 &  114.7 &  101.1 & -410.0 \\
5795111411562833152 &  312.90 & -12.76 &      - &  6.3 &  2.0 & -0.6 &   14.7 &   48.0 &  114.5 & -273.0 \\
6446682796754391936 &  338.27 & -28.03 &  -0.86 &  6.6 &  0.7 & -0.9 &   83.1 &   42.8 &  124.0 & -225.7 \\
6189105151220002816 &  311.71 &  34.25 &  -1.75 &  6.7 &  1.7 &  1.6 &  -56.3 &   34.0 &  -68.7 & -323.6 \\
6482200419648454144 &  351.91 & -37.26 &  -1.06 &  6.1 &  0.3 & -1.6 &   68.6 &   54.7 &  -94.1 & -315.0 \\
6793672002009076864 &   13.51 & -35.00 &      - &  4.6 & -0.9 & -2.6 & -126.2 &   57.2 &   63.3 & -153.6 \\
6662369212476955136 &  348.99 & -20.38 &  -1.25 &  6.3 &  0.4 & -0.7 &   69.6 &   55.7 &  -81.0 & -325.2 \\
4354802482560690048 &   12.75 &  27.77 &  -2.03 &  5.9 & -0.5 &  1.3 &  -41.8 &   15.1 & -111.0 &  -67.1 \\
6855988919456460928 &   22.31 & -30.34 &   -0.9 &  4.8 & -1.4 & -2.1 &  -95.6 &   51.8 &   73.9 & -114.2 \\
5807722092676009600 &  319.95 & -16.37 &  -0.98 &  7.1 &  0.9 & -0.4 &   79.2 &   37.2 &   92.3 & -193.9 \\
6617500174464520832 &   18.51 & -51.33 &      - &  6.1 & -0.7 & -2.7 &   25.8 &   21.0 &  -79.7 & -147.0 \\
6390694633874538496 &  316.76 & -45.53 &  -2.11 &  7.6 &  0.5 & -0.8 &  -56.5 &   42.9 &  -94.2 & -357.2 \\
6593804217121507968 &    1.38 & -58.28 &  -1.08 &  6.4 & -0.0 & -2.9 &   12.5 &   26.9 &  -85.4 & -171.7 \\
6511387028823212544 &  342.42 & -51.60 &  -2.28 &  7.0 &  0.4 & -1.5 &   33.0 &   53.7 &  -82.2 & -365.9 \\
6777598718260083584 &   10.49 & -42.97 &   -1.1 &  5.9 & -0.4 & -2.2 &   12.2 &   55.5 &   94.3 & -330.7 \\
6667233009535349888 &  349.07 & -32.38 &  -2.27 &  4.6 &  0.7 & -2.3 & -158.1 &   -6.4 &   49.9 &  -82.0 \\
6082279178445547520 &  309.20 &  12.41 &  -0.62 &  6.3 &  2.4 &  0.7 &  116.5 &   91.5 &   92.2 & -297.2 \\
6305435577383801472 &  340.62 &  35.43 &  -0.63 &  6.2 &  0.7 &  1.5 &  -63.9 &   17.1 &  129.9 & -150.9 \\
4381400665271860480 &   19.62 &  25.70 &  -0.86 &  4.6 & -1.3 &  1.9 &  124.8 &   30.3 &  120.1 & -300.1 \\
6497164021283998976 &  318.68 & -60.57 &  -1.06 &  7.7 &  0.5 & -1.2 &  -11.1 &   19.7 &   72.8 & -156.6 \\
6317919780298422272 &  354.53 &  37.65 &  -1.42 &  5.5 &  0.3 &  2.1 &  -68.2 &   72.9 & -102.1 & -421.0 \\
6443381955472092160 &  337.29 & -33.10 &  -1.56 &  5.0 &  1.4 & -2.3 &   81.2 &   58.4 &  -68.8 & -180.5 \\
6415791880133409920 &  321.64 & -27.75 &      - &  6.9 &  1.1 & -0.9 &   71.3 &   40.3 &  -90.3 & -201.0 \\
5531106613768555648 &  259.73 &  -9.26 &      - &  8.3 &  0.4 & -0.0 &   46.6 &   38.5 &  -79.3 & -300.5 \\
6003500922668306944 &  328.93 &  12.10 &  -0.65 &  4.9 &  2.0 &  0.9 &   76.8 &   88.4 & -147.6 & -274.7 \\
2597932671179367936 &   48.94 & -55.83 &      - &  8.0 & -0.2 & -0.3 &  -29.1 &   36.4 &   83.3 & -287.2 \\
6753447365539976448 &   11.45 & -24.54 &      - &  4.4 & -0.8 & -1.7 &   -2.0 &   27.7 &  168.0 & -121.1 \\
6259344638358421504 &  346.72 &  33.62 &  -1.39 &  5.5 &  0.6 &  1.9 & -105.6 &   23.1 &  -93.0 & -194.2 \\
6267173332950478336 &  354.53 &  32.41 &      - &  5.5 &  0.3 &  1.7 &   21.4 &   73.8 & -145.6 & -402.6 \\
6858853692704665856 &   28.81 & -33.91 &      - &  6.4 & -1.0 & -1.4 &   38.5 &   44.4 &  112.4 & -321.5 \\
6575958662367490816 &  354.95 & -46.37 &  -1.86 &  6.5 &  0.1 & -1.7 &   14.1 &   29.2 &   92.3 & -189.0 \\
6649706450595363712 &  339.06 & -20.93 &  -1.45 &  5.5 &  1.1 & -1.1 & -170.2 &   21.2 &   66.2 & -294.3 \\
5796181476896205696 &  309.48 & -13.05 &  -1.15 &  4.8 &  4.1 & -1.2 &   19.7 &   51.8 & -100.1 & -168.0 \\
6262591290040852096 &  354.40 &  28.33 &  -1.66 &  5.8 &  0.2 &  1.3 &   17.1 &   59.3 &   88.7 & -340.8 \\
6491444052558679424 &  325.39 & -52.77 &      - &  7.9 &  0.2 & -0.5 &   32.5 &   15.4 &  -76.8 & -113.3 \\
6835943340714822016 &   34.27 & -40.98 &      - &  6.3 & -1.3 & -1.9 &   74.2 &   17.0 &  135.5 & -202.2 \\
6695050172681806080 &    5.27 & -34.07 &      - &  6.8 & -0.1 & -1.0 &   66.5 &   38.8 &  -66.1 & -271.2 \\
6484804307998121600 &  357.52 & -41.88 &  -1.81 &  5.2 &  0.1 & -2.7 &  104.5 &   89.2 &  -63.5 & -451.4 \\
6426826888067362944 &  327.79 & -31.15 &      - &  4.7 &  2.2 & -2.5 &   55.2 &   44.4 & -103.0 &  -85.6 \\
6135575638078558080 &  303.84 &  18.99 &  -0.86 &  7.0 &  1.7 &  0.7 &   97.8 &   65.1 &   90.2 & -286.9 \\
6849425139977703040 &   19.65 & -31.12 &  -1.44 &  4.3 & -1.4 & -2.5 &   63.0 &   31.0 & -101.8 & -220.8 \\
6676827996537504896 &  357.93 & -39.33 &  -0.54 &  5.0 &  0.1 & -2.6 &   48.8 &   12.5 &  -92.9 &  -56.5 \\
3637344787222734592 &  323.91 &  58.37 &  -0.92 &  7.0 &  0.8 &  2.4 &  -36.7 &   29.1 &  102.9 & -235.6 \\
6531487956804986752 &  333.30 & -67.53 &   -0.8 &  7.2 &  0.5 & -2.7 &  -43.0 &   22.4 &   89.1 & -182.9 \\
6658598265547671936 &  346.95 & -25.55 &      - &  5.3 &  0.7 & -1.4 &  103.8 &   36.5 &  -86.6 & -121.4 \\
6848010210249401856 &   16.83 & -30.86 &  -0.61 &  5.7 & -0.8 & -1.5 &  -79.5 &   71.6 & -141.5 & -347.8 \\
5819645609098141824 &  318.78 & -12.46 &  -1.43 &  5.6 &  2.3 & -0.7 &   24.9 &   61.4 & -101.9 & -289.4 \\
6478853059576336768 &  349.20 & -43.11 &  -1.67 &  7.2 &  0.2 & -0.9 &   53.5 &   19.4 &  109.4 & -130.5 \\
6582323258440558848 &    1.65 & -44.70 &  -2.34 &  6.1 & -0.1 & -2.1 &   68.3 &   36.9 &  -69.0 & -228.5 \\
6260477753814244608 &  351.60 &  28.36 &      - &  6.6 &  0.2 &  0.9 &  -45.9 &   36.4 &  -99.3 & -251.9 \\
6814677347020224000 &   24.96 & -47.69 &  -1.38 &  7.3 & -0.4 & -1.1 &  -14.0 &   49.7 & -102.2 & -355.3 \\
6704697528226119552 &  348.94 & -18.76 &      - &  4.8 &  0.7 & -1.1 &  122.8 &   76.9 &  154.0 & -290.2 \\
6269052359662541440 &  324.22 &  29.93 &      - &  5.6 &  1.8 &  1.8 & -149.4 &    2.8 &  -76.4 & -290.8 \\
6463190104122275968 &  342.04 & -42.00 &      - &  6.0 &  0.7 & -2.1 &   -4.8 &   17.8 &  -95.5 & -110.4 \\
5786628095241089408 &  308.11 & -15.41 &  -1.67 &  7.7 &  0.6 & -0.2 &  -17.8 &   39.5 &  -99.2 & -315.7 \\
5913106227764290176 &  333.24 & -13.38 &      - &  6.1 &  1.0 & -0.5 &   24.0 &   34.1 & -108.8 & -184.3 \\
6278040695420767104 &  331.57 &  32.66 &      - &  5.2 &  1.6 &  2.2 &  -12.0 &   54.6 &  118.7 & -302.3 \\
4378858147711990400 &   15.26 &  26.08 &  -1.19 &  4.3 & -1.1 &  2.0 & -109.2 &   69.4 &  -85.7 & -186.8 \\
6512412907532485376 &  334.55 & -57.07 &  -2.37 &  7.4 &  0.4 & -1.3 &   17.5 &   32.0 &  -81.7 & -230.8 \\
4403212776946445696 &    5.26 &  38.55 &  -1.91 &  6.1 & -0.2 &  1.7 & -102.8 &   30.7 & -101.4 & -165.6 \\
5912911678614539264 &  333.42 & -14.12 &  -0.97 &  4.9 &  1.7 & -0.9 &  -54.1 &   32.9 & -127.7 & -250.1 \\
6768555926615706624 &   17.19 & -21.63 &  -0.66 &  6.2 & -0.6 & -0.8 &  103.7 &   24.0 &  -98.7 & -213.1 \\
5915043425508255488 &  330.27 & -10.37 &      - &  5.5 &  1.6 & -0.6 &  -67.7 &    3.1 & -125.0 & -122.8 \\
6421895681498235008 &  327.66 & -27.84 &  -1.57 &  3.5 &  3.0 & -2.9 &   17.7 &   20.6 & -117.6 &  -19.0 \\
6558374658235864960 &  343.17 & -47.90 &      - &  7.1 &  0.3 & -1.2 &  -93.7 &   42.1 &   67.8 & -330.7 \\
5789859839093874048 &  306.39 & -13.48 &      - &  6.2 &  2.7 & -0.8 &   44.4 &   31.6 &  108.5 &  -74.6 \\
6147829420291613312 &  293.98 &  19.21 &  -1.66 &  7.2 &  2.3 &  0.9 &  -40.7 &    7.0 &  -85.0 & -145.6 \\
5809476294773601408 &  321.46 & -12.93 &      - &  5.3 &  2.3 & -0.8 & -114.4 &  -13.2 &   86.2 & -190.0 \\
6473910342493757824 &  347.11 & -34.07 &  -0.54 &  7.2 &  0.2 & -0.7 &  -77.4 &   25.4 & -110.9 & -200.1 \\
\hline
\end{longtable}
\end{center}

\begin{center}
\begin{longtable}{ccccccccccc}
\caption{GR-2, 3.62 $\sigma$, 34 stars}\\
\hline
\hline
source-id & l &      b &  [Fe/H] &     x &    y &     z &      U &      V &      W &     $L_z$     \\
& (deg) & (deg) & (dex) & (kpc) & (kpc) & (kpc) & (km s$^{-1}$) & (km s$^{-1}$) & (km s$^{-1}$) & (km s$^{-1}$ kpc) \\
\hline\\
6449845198355290240 &  330.85 & -39.03 &  -1.31 &  7.2 &  0.5 & -0.9 &  -43.8 & -107.5 &  -41.7 &  754.9 \\
6258852675626795392 &  349.00 &  32.38 &  -1.25 &  5.4 &  0.5 &  1.8 &  132.6 & -113.8 &  -88.8 &  690.0 \\
6317052506146717312 &  357.69 &  34.39 &  -1.07 &  6.2 &  0.1 &  1.4 &   17.4 &  -74.9 &   43.9 &  469.4 \\
6004194645781493888 &  329.88 &  13.49 &  -0.99 &  5.1 &  1.8 &  0.9 &  137.6 &  -78.2 &   63.5 &  646.3 \\
5789526515270303872 &  306.13 & -14.70 &  -0.81 &  7.7 &  0.7 & -0.2 &   10.3 &  -66.6 &  -57.1 &  520.3 \\
4348288219771571200 &    2.98 &  30.15 &  -0.84 &  5.6 & -0.1 &  1.5 &  -41.9 & -124.0 &  -46.4 &  702.6 \\
6158877588002891648 &  302.54 &  28.80 &  -1.28 &  6.4 &  2.8 &  1.8 &   35.2 &  -77.7 &   37.2 &  597.4 \\
5790486663802736896 &  308.13 & -12.20 &   -1.0 &  5.6 &  3.3 & -0.9 &   91.8 &  -72.6 &   56.7 &  711.0 \\
6218157791784797184 &  331.26 &  26.34 &  -0.73 &  6.3 &  1.1 &  1.1 &  119.1 &  -75.2 &   32.8 &  597.6 \\
6114741713797552384 &  317.33 &  23.45 &  -1.25 &  5.1 &  2.9 &  1.9 &   47.9 &  -70.9 &   70.7 &  498.4 \\
6665690974540896640 &  345.53 & -31.25 &      - &  5.3 &  0.8 & -1.8 &  -55.8 & -109.9 &   56.7 &  535.9 \\
6251112216848937984 &  342.22 &  27.38 &  -0.67 &  5.3 &  0.9 &  1.6 &   94.3 &  -73.4 &   45.3 &  474.8 \\
4404899977538537856 &    7.05 &  34.56 &  -0.76 &  4.7 & -0.4 &  2.5 &  -46.1 & -128.3 &  -53.8 &  617.9 \\
6074173720472593152 &  303.28 &   9.29 &  -0.62 &  6.7 &  2.3 &  0.5 &  -25.3 &  -97.7 &  -64.1 &  597.7 \\
6511309066576795136 &  342.42 & -52.23 &  -2.11 &  6.1 &  0.7 & -2.8 &   38.8 &  -90.6 &   25.3 &  580.4 \\
4342924973847237760 &  358.88 &  29.57 &  -1.23 &  4.9 &  0.1 &  1.9 &  101.9 & -144.6 &    0.7 &  710.4 \\
3506830356119273728 &  314.08 &  42.52 &      - &  6.3 &  2.0 &  2.6 &   12.1 & -104.7 &   10.2 &  680.0 \\
6788387646043341824 &   15.38 & -40.50 &  -0.63 &  6.4 & -0.5 & -1.6 &  -10.6 &  -91.0 &  -39.8 &  587.8 \\
6435386925388937472 &  330.59 & -25.94 &      - &  4.1 &  2.3 & -2.3 &  -27.2 & -130.3 & -115.4 &  465.8 \\
6313449367881650688 &  344.81 &  41.32 &      - &  5.1 &  0.8 &  2.8 &   71.2 &  -94.0 &   -3.8 &  540.1 \\
6646407636533587712 &  346.56 & -27.10 &  -1.02 &  7.2 &  0.2 & -0.5 &  -47.5 &  -74.1 &   49.9 &  518.8 \\
6309259232147407616 &  348.38 &  37.29 &      - &  6.3 &  0.4 &  1.5 &   16.7 & -109.8 &  -56.2 &  697.6 \\
6589763718048296064 &    9.28 & -48.73 &      - &  6.5 & -0.3 & -2.0 &    4.3 &  -76.4 &  -39.1 &  492.9 \\
5810135417625152128 &  323.29 & -19.57 &  -0.51 &  6.9 &  0.9 & -0.5 &   38.3 &  -93.2 &   50.4 &  683.5 \\
5818135185049009664 &  327.52 & -12.48 &      - &  5.5 &  1.7 & -0.7 &  150.8 &  -72.6 &  -74.1 &  657.8 \\
6480211231315277056 &  352.33 & -42.67 &  -0.67 &  6.0 &  0.3 & -2.0 &  -82.2 &  -98.0 &   25.0 &  567.5 \\
4357650217678900736 &   11.60 &  29.84 &      - &  6.2 & -0.4 &  1.2 &   98.1 &  -97.7 &   11.8 &  568.9 \\
6428120360420225024 &  329.39 & -28.65 &  -1.15 &  4.5 &  2.2 & -2.4 &  -54.0 & -152.4 &  -72.1 &  558.7 \\
6240366689710897408 &  348.61 &  23.95 &  -0.19 &  4.3 &  0.8 &  1.8 &  -37.1 & -162.6 &  -36.1 &  668.5 \\
6632745826600895232 &  334.74 & -22.72 &      - &  7.3 &  0.4 & -0.4 &  -74.9 &  -83.6 &   46.0 &  578.8 \\
6713848351184784640 &  355.03 & -21.46 &  -0.55 &  5.3 &  0.2 & -1.1 &  124.8 & -105.3 &  113.3 &  591.8 \\
5768695992600783744 &  309.10 & -23.19 &  -1.03 &  5.5 &  3.3 & -1.8 &  116.5 &  -28.2 &  -24.2 &  539.5 \\
5819460581907006976 &  318.99 & -12.95 &  -0.54 &  4.0 &  3.7 & -1.3 &  118.0 &  -91.8 &  -86.3 &  798.3 \\
6260416898417038336 &  350.39 &  28.86 &  -0.53 &  4.9 &  0.6 &  1.9 &  -30.0 & -172.5 &   18.5 &  828.0 \\
		\hline
	\end{longtable}
\end{center}

\end{document}